\documentclass[12pt]{iopart}
\usepackage[utf8]{inputenc}
\usepackage{setspace}
\usepackage[square,sort&compress,numbers]{natbib}

\usepackage{graphicx}%
\graphicspath{ {figures/} }
\usepackage{lipsum} %
\usepackage{wrapfig}%
\usepackage{ifthen}%
\usepackage{color} %
\usepackage{bm} %
\usepackage{multirow}%
\usepackage{bbold} %
\usepackage{csquotes}

\def\QP{\textrm{QP}}

\begin{document}

\title[Computation of  $\langle \hat{S}^2 \rangle$ for the Spin-Flip Bethe-Salpeter Equation]{Computation of the expectation value of the spin operator $\hat{S}^2$ for the Spin-Flip Bethe-Salpeter Equation}

\author{B A Barker$^{1,2}$, A Seshappan$^2$ and D A Strubbe$^2$}
\address{$^1$ Department of Materials Science and Engineering, University of California,
  Merced, CA 95343, USA}
\address{$^2$ Department of Physics, University of California,
  Merced, CA 95343, USA}

\begin{abstract}
Spin-flip methods applied to excited-state approaches like the Bethe-Salpeter Equation allow access to the excitation energies of open-shell systems, such as molecules and defects in solids. The eigenstates of these solutions, however, are generally not eigenstates of the spin operator $\hat{S}^2$. Even for simple cases where the excitation vector is expected to be, for example, a triplet state, the value of $\langle \hat{S}^2 \rangle$ may be found to differ from 2.00; this difference is called ``spin contamination.'' The expectation values $\langle \hat{S}^2 \rangle$ must be computed for each excitation vector, to assist with the characterization of the particular excitation and to determine the amount of spin contamination of the state. Our aim is to provide for the first time in the spin-flip methods literature a comprehensive resource on the derivation of the formulas for $\langle \hat{S}^2 \rangle$ as well as its computational implementation. After a brief discussion of the theory of the Spin-Flip Bethe-Salpeter Equation and some examples further illustrating the need for calculating $\langle \hat{S}^2 \rangle$, we present the derivation for the general equation for computing $\langle \hat{S}^2 \rangle$ with the eigenvectors from an SF-BSE calculation, how it is implemented in a Python script, and timing information on how this calculation scales with the size of the SF-BSE Hamiltonian.
\end{abstract}

\maketitle

\section{Introduction}
\label{sec:introduction}

Spin-flip methods have been introduced in the electronic structure literature beginning in 2001 \cite{krylov2001size}, developed for post-Hartree-Fock methods \cite{krylov2006spin} and time-dependent density-functional theory (TDDFT) \cite{shao2003spin}. These approaches describe states of an open shell system via spin-flipping excitations from a high-spin reference state that can be accurately described by Hartree-Fock or density-functional theory (DFT). The approach was initially used for open-shell molecules\cite{krylov2001size}, but recently SF-TDDFT has been successfully applied to defects in solids \cite{jin2023excited}. An important aspect of these methods is the evaluation of matrix elements of the spin operator $\hat{S}^2$ to assist with analysis of the computed results, and assess the presence of spin contamination \cite{Sonnenberg} which is notorious in such approaches \cite{lee2018eliminating}. While Ref. \cite{jin2023excited} does not comment on the spin contamination of their computed results for energy levels of defected solids, analysis of the expectation value of $S^2$ for these states would be useful. Ref. \cite{li2011spin} presents in its appendix the necessary equation to calculate $\hat{S}^2$ for spin-flip excitations. However, a detailed derivation is missing and the authors are not aware of its presence elsewhere in the literature. Obtaining the end result, beginning from the so-called ``super-operator'' approach, is not trivial. In this article, we provide a thorough derivation of this equation, to complement the derivation of the spin-conserving version presented in Ref. \cite{ipatov2009excited}. We include many steps in working out the normal ordering of the creation and annihilation operators in the nested commutators to assist any readers interested in working through the results presented in Ref. \cite{li2011spin} and \cite{ipatov2009excited}, or applying the ``super-operator'' approach to previously unconsidered transitions. We additionally include information about the practical implementation for calculating matrix elements of $\hat{S}^2$ with spin-flip excited states, using examples first considered in Ref. \cite{barker2022spin} to illustrate results, considerations, and time performance of the relatively straightforward serial Python scripts.

We briefly review some of the procedure to perform a spin-flip (SF) calculation, especially as applied to the Bethe-Salpeter Equation (BSE) \cite{Strinati,rohlfing2000electron}. The spin-flip BSE (SF-BSE) approach for open-shell systems has been developed independently by Monino and Loos \cite{monino2021spin} and by Barker and Strubbe \cite{barker2022spin}. Monino and Loos applied the method to atoms and molecules. Barker and Strubbe applied the method to molecules but also quantum defects in solids, adding to the available approaches for tackling these challenging systems \cite{Gali}. (That work also pointed out the theoretical problems with using conventional GW calculations with SF-BSE.) In the procedure for an SF-BSE calculation, first the orbitals (both occupied and unoccupied) of the high-spin reference state $|\textrm{H.S. Ref}\rangle$ are computed with DFT methods. The screened Coulomb interaction $W$ is computed using these Kohn-Sham (KS) orbitals and energies as in GW calculations \cite{hybertsen1986electron,BerkeleyGW}. The BSE Kernel $K$ is constructed using only transitions from occupied up-spin states to unoccupied down-spin states, and the SF-BSE Hamiltonian is then diagonalized. The eigenvalue equation is
\begin{equation}
\left(E^{\QP}_{\bar{a}\downarrow} - E^{\QP}_{i\uparrow} \right)A^{I}_{i\uparrow,\bar{a}\downarrow}\, + \sum_{j,\bar{b}} K_{j\uparrow \bar{b}\downarrow,i\uparrow \bar{a}\downarrow} A^{I}_{i\uparrow,\bar{a}\downarrow} = \Omega^I A^{I}_{i\uparrow,\bar{a}\downarrow} \, ,
\end{equation}
with quasiparticle energies $E^{\QP}$ for unoccupied spin-down states $\bar{a}$ and occupied spin-up $i$, excitation energy $\Omega^I$, and excitation eigenvector components $A^I_{i\uparrow,\bar{a}\downarrow}$. (The superscript ``$I$'' is used as the index for some particular excitation.)  The eigenvalues $\Omega^I$ can be used to calculate observables such as (``vertical,'' i.e. ground-state geometry) optical transition energies by taking differences of these eigenvalues, after identifying the character of the eigenstates $|\Psi^I \rangle$. In this way, optical transition energies of open-shell systems, such as defected solids, may be computed without needing to determine the complete theory of BSE as applied to open-shell systems.

The eigenstates $|\Psi^I \rangle$ are linear combinations of ``target states,'' which are the states generated from a spin-flip excitation of the high-spin reference state, given by
\begin{equation}
|\Psi^I\rangle = \sum_{i,\bar{a}} A^I_{i,\bar{a}} \bar{a}^{\dagger}_{\downarrow}i_{\uparrow} | N, 0 \rangle \, ,
\end{equation}
where $|N,0\rangle$ is the high-spin reference state, $i_{\uparrow}$ is the annihilation operator for the up-spin orbital labeled as ``$i$'' in the reference determinant, $\bar{a}^{\dagger}_{\downarrow}$ is the creation operator for the down-spin orbital labeled as ``$\bar{a}$'' in the reference determinant, and $A^I_{i,\bar{a}}$ is the amplitude for the particular spin flip transition. (The convention with the use of the overbar is elucidated in Sec. \ref{sec:necessity}.) We note that are exclusively considering excited states within the Tamm-Dancoff approximation \cite{fetter1971quantum}, where our excited states $|\Psi^I \rangle$ do not have contributions from de-excitations; Refs. \cite{ipatov2009excited,myneni2017calculation,li2011spin} consider non-Tamm-Dancoff states. 
The target states $\bar{a}^{\dagger}_{\downarrow}i_{\uparrow} | N, 0 \rangle$, or ``$|i_{\uparrow},\bar{a}_{\downarrow}\rangle$'', as with the basis set of single-particle transitions of ordinary BSE, may be characterized by the pair of states involved in the single-particle transition. 
The space of target states $|i\uparrow,\bar{a}\downarrow\rangle$ form the basis set to describe multiple configurations associated with the ground and excited states of open-shell systems.

This article is organized as follows. In Section \ref{sec:previous}, we discuss details regarding our previous SF-BSE calculation results for the test systems, ethylene under torsion and the NV$^{-}$ center in diamond.  In Section \ref{sec:necessity}, we discuss why the explicit calculation of $\langle \hat{S}^2 \rangle$ is necessary. In Section \ref{sec:contamination}, we discuss sources of deviation of the computed value of $\langle \hat{S}^2 \rangle$ from naive expectation. In Section \ref{sec:derivation}, we provide the derivation of the formula for $\langle \hat{S}^2 \rangle$ for spin-flip excitations. In Section \ref{sec:example}, we use a simple example to illustrate how to use that formula. Finally, in Section \ref{sec:timing}, we discuss the scripts and their performance in calculating $\langle \hat{S}^2 \rangle$.

\section{Summary of example calculations with SF-BSE}
\label{sec:previous}

The authors have previously calculated the ground and excited states for interesting open-shell systems in Ref. \cite{barker2022spin}, using an implementation with a modified version of the BerkeleyGW code \cite{BerkeleyGW}, revision 7294, roughly equivalent to public release version 2.1. Among these, first, is the ethylene molecule (C$_2$H$_4$) under torsion (from $0^\circ$ to $90^\circ$). The second is the NV$^{-}$ center in diamond, a defect in crystalline diamond-structure carbon in which one carbon atom is substituted with nitrogen (``N''), a neighboring atom is left vacant (``V''), and the defect bears an overall negative charge
(``$^{-}$''). Here we summarize the key states and computational parameters for these previous calculations, which we further analyze in this work.

The ethylene molecule under torsion is a conventional test system for computational electronic structure approaches that allow for the description of open-shell states \cite{shao2003spin}. Under zero torsion, the ethylene molecule is a closed-shell singlet. At $90^\circ$ of torsion (about the carbon-carbon double-bond), however, the ground-state is an open-shell triplet. There are four many-body electronic states whose energies give potential surfaces with respect to the torsion angle. The ``N'' state is the lowest energy singlet (the ground state, under zero torsion). The ``T'' state is the lowest energy triplet (the ground state, at $90^\circ$ of torsion). The ``V'' and ``Z'' states are higher-energy singlets. 

The DFT calculation in Ref. \cite{barker2022spin} of the ethylene orbitals was performed with the DFT code Octopus \cite{octopus2015,octopus2020}, version 8.4. We used  the Optimized Norm-Conserving Vanderbilt pseudopotentials \cite{hamann2013optimized} from the Pseudo-Dojo pseudopotential database \cite{van2018pseudodojo}, version 0.2, with the PBE exchange-correlation \cite{perdew1996generalized}. The relaxed atomic coordinates (at $0^{\circ}$) were calculated with a 0.115~{\AA} real-space grid spacing, roughly equivalent to a 115~{Ry} planewave wavefunction cutoff, in a box with edge-length 12~{\AA}. A smaller box size was used in the subsequent calculations, which contains 99\% of the charge density for both the ethylene molecule with no torsion and $90^{\circ}$ of torsion. 
The ground- and excited-state energies were calculated explicitly at torsions of 0, 5, 10, 15, 30, 45, 60, 75, 80, 85, and 90$^{\circ}$. For DFT input to SF-BSE calculations, we used a less stringent real-space grid spacing for both the wavefunctions and density of 0.18~{\AA} , equivalent to about 85 Ry planewave cutoff for the wavefunctions. We obtained the $M_S=1$ reference state by constraining occupations. Based on \cite{van2015gw}, we used 860 empty states and 24~{Ry} for the calculation of the screened Coulomb interaction. For the SF-BSE Hamiltonian, 5 occupied spin-up (the maximum allowed) and 55 unoccupied spin-down orbitals were used.

\begin{figure}
\centering
\includegraphics[scale=0.8]{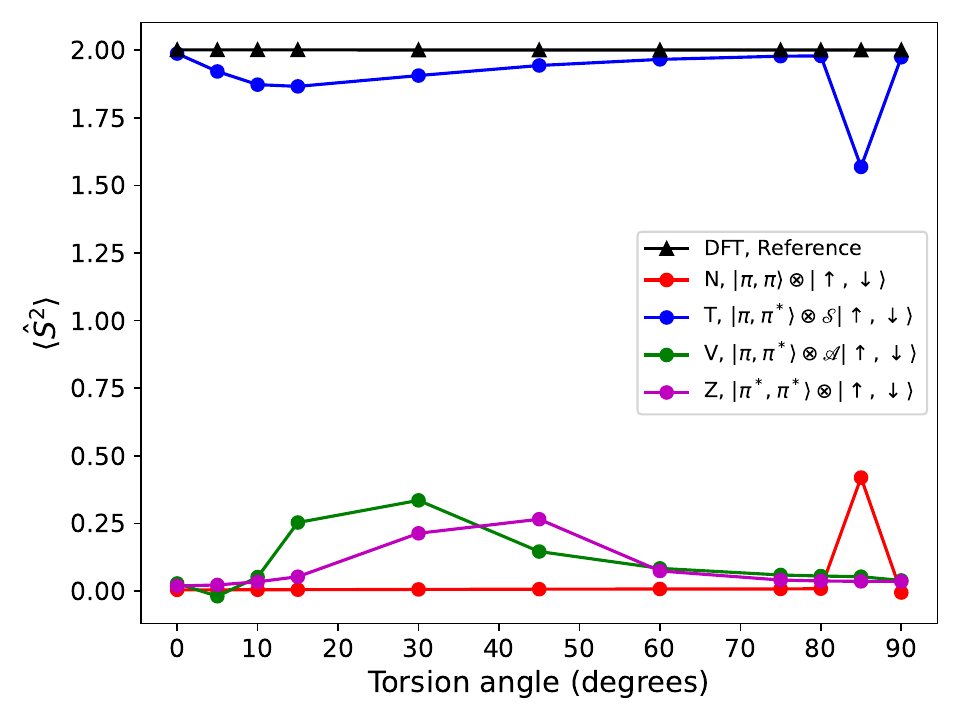}
\caption{The computed values of $\langle \hat{S}^2 \rangle$ for the high-spin reference state and the N, T, V, and Z states of ethylene within SF-BSE. T is a triplet while the others are singlets. The spin of the many-body states is treated with a symmetric (``$\mathcal{S}$'') or antisymmetric (``$\mathcal{A}$'') combination of the individual spins. }
\label{fig:spin_contamination}
\end{figure}

For the ethylene molecule under torsion, we then calculate $\langle \hat{S}^2 \rangle$ for the SF-BSE excitation vectors for the N, T, V, and Z states. The results are shown in Fig. \ref{fig:spin_contamination}. We note that a rather high amount of contamination occurs at $85^\circ$, when the singlet N and triplet T states cross in energy, with the triplet now lower \cite{barker2022spin}. Otherwise the N and T states have relatively low contamination. The V and Z states, however, show larger contamination, and at lower angles. The Z state, for instance, is difficult to discern from other excitations at these angles and this phenomenon can be understood as a measure of the incompleteness of the basis of transitions used to describe this state.

The NV$^{-}$ center in diamond is a well-studied open-shell defect in a crystalline solid. Within its bulk band gap are so-called in-gap single-particle states labeled ``$v$'' and ``${e_x,e_y}$'', the latter being doubly-degenerate. (See, for instance, Ref. \cite{choi2012mechanism}.) Its ground state is a triplet, with the $S = 1, M_S = 1$ state having its up-spin $v$, $e_x$, and $e_y$ orbitals occupied and only its down-spin $\bar{v}$ orbital occupied. (For SF-BSE calculations, this is our high-spin reference state.) Its three-fold degenerate triplet ground state is labeled $^3A_2$, and its optical excitations, therefore, are not accessible by conventional GW/BSE. We compute its transition energies to the singlet states $^1E$, $^1A_1$ and the triplet state $^3E$ in Ref. \cite{barker2022spin}. Again, we compute the bands in DFT with the code Octopus \cite{octopus2015,octopus2020}. The defect is placed in a 2$\times$2$\times$2 supercell, with 63 atoms (62 carbon, 1 nitrogen), since the orbitals of the in-gap states are well-localized \cite{choi2012mechanism}. The real-space grid spacing is 0.34 Bohr, again, equivalent to an 85 Ry planewave cutoff. The pseudopotentials are also Optimized Norm-Conserving Vanderbilt pseudopotentials \cite{hamann2013optimized} from the Pseudo-Dojo pseudopotential database \cite{van2018pseudodojo} with the PBE exchange-correlation\cite{perdew1996generalized}. The dielectric matrix is calculated with 300 empty states and a cutoff of 12~{Ry}, consistent with the choice of parameters in Ref.  \cite{lischner2012first}. The SF-BSE Hamiltonian is constructed from 12 occupied and 11 unoccupied orbitals.

\section{Necessity for the explicit calculation of $\langle \hat{S}^2 \rangle$}
\label{sec:necessity}

With the eigenvectors and eigenvalues from a successful SF-BSE calculation, one may wonder why the additional step for calculating $\langle \hat{S}^2 \rangle$ is necessary. With appropriate mapping of the indices of the up-spin hole and down-spin electron in the excitation eigenvectors to electronic configurations, one may suppose that signs of the coefficients read out explicitly from the eigenvector data give us sufficient information to determine the spin symmetry of the state.

Let us consider an example: the excitations as computed from SF-BSE for ethylene, at zero torsion. The single-particle orbitals that describe the HOMO and LUMO (as a minimal basis set) are $|\pi\uparrow\rangle$, $|\pi^*\uparrow\rangle$, $|\bar{\pi}\downarrow\rangle$, and $|\bar{\pi}^*\downarrow\rangle$. The use of the overbars for the down-spin states denotes the different spatial wavefunction, compared to the up-spin state, as occurs in a spin-polarized or spin-unrestricted approach. The high-spin reference state is the triplet $|\pi\uparrow,\pi^*\uparrow\rangle$. The excitation vectors of interest describe the following states\cite{merer1969ultraviolet}, with approximately the following numerical coefficients (to two decimals) and signs\cite{barker2022spin}:
\begin{eqnarray}
& |N\rangle  & = 0.98 |\pi\uparrow,\bar{\pi}\downarrow\rangle + 0.20 |\pi^*\uparrow,\bar{\pi}^*\downarrow\rangle \,, \\
& |T\rangle & = 0.71 |\pi\uparrow,\bar{\pi}^*\downarrow\rangle + 0.71 |\pi^*\uparrow,\bar{\pi}\downarrow\rangle \, ,\\
& |S\rangle & = 0.71 |\pi\uparrow,\bar{\pi}^*\downarrow\rangle - 0.71 |\pi^*\uparrow,\bar{\pi}\downarrow\rangle \, ,\\
& |Z\rangle & = -0.20 |\pi\uparrow,\bar{\pi}\downarrow\rangle + 0.98 |\pi^*\uparrow,\bar{\pi}^*\downarrow\rangle \,
\end{eqnarray}

\begin{figure}
\centering
\includegraphics[scale=0.4]{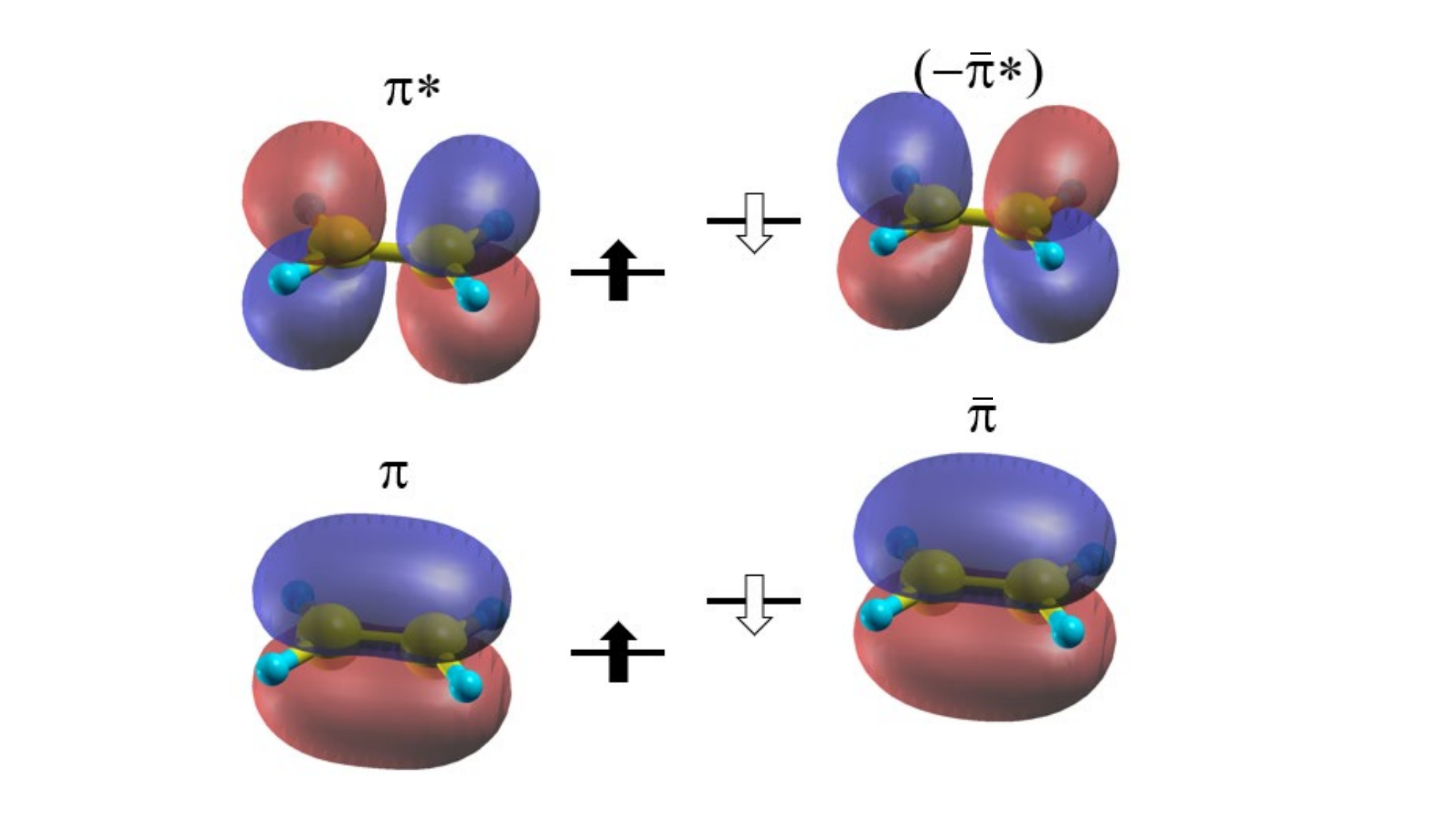}
\caption{The frontier orbitals of ethylene under zero torsion, in the high-spin reference state (up-spin orbitals occupied), plotted with XCrysDen \cite{xcrysden}. The down-spin $\bar{\pi}^*$ orbital in this illustration has a reversal of phase relative to the up-spin orbital.}
\label{fig:phase_flip}
\end{figure}

In a ``spin-restricted'' calculation, the eigenstates from a spin-unpolarized calculation (i.e., disregarding spin entirely), are doubled and placed into two spin channels, with the occupations made consistent with the magnetization of the system, {\it post facto}. The orbitals in both spin channels are necessarily the same, including the energies and phase information. However, in spin-polarized calculations, in which the different spin channels have independent Hamiltonians, the orbitals with the same band or orbital index but different spins generally have different real-space wavefunctions and energies. (These are called ``spin-unrestricted calculations''). In a spin-unrestricted calculation, therefore, there will likely be some phase rotation of the orbitals in the down-spin channel relative to the up-spin channel. If, for example, $|\bar{\pi}^*\downarrow\rangle$ orbital has a change relative to the $|\pi^*\uparrow\rangle$ orbital (see Fig. \ref{fig:phase_flip}), a naive reading of the excitation vector coefficients would swap the interpretation of the $|T\rangle$ and $|S\rangle$ states.

In general, the introduction of a phase in an orbital $| n\rangle \rightarrow e^{i\phi}|n\rangle$ is accompanied with a phase shift in the numerical value of the eigenvector coefficient $A^I_{mn} \rightarrow e^{-i\phi} A^I_{mn}$. That is, the wavefunction describing the excitation $I$ computed within SF-BSE $|\Psi^I\rangle = \sum_{i\bar{a}} A^I_{i\bar{a}} | i\bar{a}\rangle$ is invariant upon change of phase of the orbitals (except for a possible arbitrary overall phase). However, the values of the particular coefficients of the eigenvector $A^I_{i\bar{a}}$ do depend on the phases of the orbitals, and, as we have seen above in the simple example, can obfuscate the interpretation of the state $|\Psi^I\rangle$.

Additionally, the particular states one is interested in describing may be most quickly identified by their $\langle \hat{S}^2 \rangle$ values. For instance, the $Z$ state of ethylene, under zero torsion, is the tenth excited state, with several maximally contaminated (i.e., equal parts singlet and triplet) uninteresting states intermediate in energy between it and the lower-energy excited singlet $V$ state. While this state at zero torsion may be identified by reading the coefficients for several excitation eigenvectors, at torsion angles greater for zero, this laborious brute-force approach will not be productive. In fact, the energy ordering of the $Z$ state changes as a function of torsion angle, ultimately becoming at $90^\circ$ of torsion the fourth excitation as ordered by energy.

\section{Spin Contamination}
\label{sec:contamination}

The eigenvectors computed from SF-BSE may yield an expectation value of $\hat{S}^2$ that differs from proper values of $S\left(S+1\right)$ such as 0 (singlet, $S = 0$) or 2 (triplet $S = 1$). One source of spin contamination that even appears in the high-spin reference state before any spin-flip excitation is applied is related to the difference in the spatial wavefunctions for the different spin channels from spin-polarized DFT \cite{baker1993spin}. In the examples of ethylene under torsion, or the NV$^{-}$, this causes deviations of 0.01 or less. A larger source of spin contamination in SF-BSE comes from the absence of particular transitions from the set of target states that are required to form an eigenvector of $\hat{S}^2$ \cite{lee2018eliminating}. For example, if the complete state is $\frac{1}{\sqrt{2}}\left(|r\uparrow,\bar{p}\downarrow\rangle + |p\uparrow,\bar{r}\downarrow\rangle \right)$, sometimes one of these requisite target transitions may be inaccessible from a single spin flip from the up-spin channel to the down-spin channel. This leads to maximally-contaminated states, where the computed $\langle \hat{S}^2 \rangle \approx 1$. 

\section{Derivation of $\langle \hat{S}^2 \rangle$ for spin-flip transitions}
\label{sec:derivation}
Readers interested only in the final result should refer to Eq. \ref{eq:delta_s2}. The detailed work that follows may assist readers in following the work in, for example, Refs. \cite{li2011spin,myneni2017calculation,ipatov2009excited}.

A spin-polarized system will have $N$ total electrons, with $N_{\uparrow}$ electrons in the spin-up channel and $N_{\downarrow}$ electrons in the spin-down channel. We use the convention that our high-spin reference state (computed from DFT) is polarized such that $N_{\uparrow} > N_{\downarrow}$, and the transitions calculated from the (spin-flip) Bethe-Salpeter Equation remove an electron from an occupied spin-up state and place it in a previously unoccupied spin-down state. The number of (necessarily spin-down) unoccupied states used in the calculation is $N_{\textrm{unocc}}$. The  set of all of the orbitals we consider in the spin-flip calculations are the up-spin KS orbitals $\phi_m$ and down-spin orbitals $\bar{\phi}_n$, with $m \in \{ 1,\ldots,N_{\uparrow}\}$ and $n \in \{ 1,\ldots,N_{\downarrow},N_{\downarrow}+1,\ldots,N_{\downarrow}+N_{\textrm{unocc}}\}$. The orbitals $n \in \{ 1,\ldots,N_{\downarrow}\}$ are always occupied, since we do not flip occupied down-spin electrons to unoccupied up-spin orbitals.  For a particular excitation, one of the spin-up orbitals $\phi_i$ with $i \in \{ 1,\ldots,N_{\uparrow}\}$  becomes unoccupied  while one of the orbitals $\bar{\phi}_a$ with $a \in \{N_{\downarrow}+1,\ldots,N_{\downarrow}+N_{\textrm{unocc}}\}$ becomes occupied.

\begin{figure}
\centering
\includegraphics[scale=0.6]{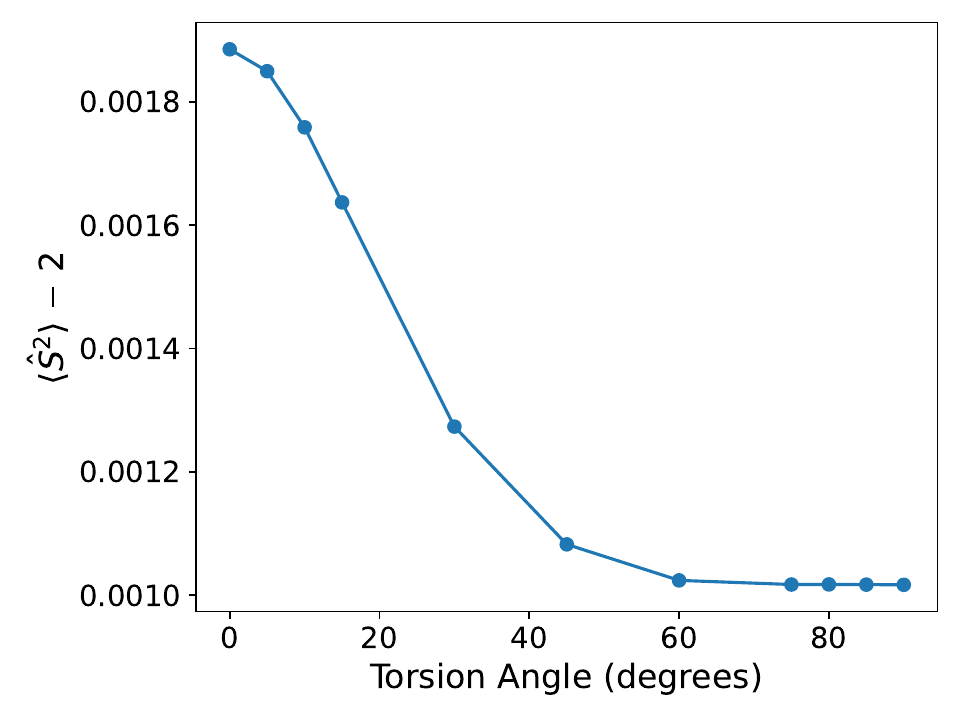}
\caption{The computed $\langle \hat{S}^2 \rangle_0$ for ethylene for torsion angles from $0^\circ$ to $90^\circ$. The High-Spin Reference state is a triplet, so the values are reported as deviations from the expected $S(S+1) = 2$. }
\label{fig:ethylene_ref_s2}
\end{figure}

We first compute $\langle \hat{S}^2 \rangle_0$ for the high-spin reference state, $|N,0\rangle$, via the following equation from the well-known L{\"o}wdin Formula  
\cite{lowdin1955quantum}:
\begin{equation}
\label{eq:lowdin}
\langle \hat{S}^2  \rangle_0 = \langle N, 0 | \hat{S}^2 | N, 0 \rangle = \left(\frac{N_{\uparrow} - N_{\downarrow}}{2} \right)\left(\frac{N_{\uparrow} - N_{\downarrow}}{2} + 1 \right) + N_{\downarrow} - \sum_{i,\bar{j}} | \langle i | \bar{j} \rangle |^2 .
\end{equation}
Using the L{\"o}wdin Formula, which is valid for a many-body wavefunction that is a Slater determinant as in Hartree-Fock, implies approximating $\langle \hat{S}^2 \rangle_0$ via a Slater determinant formed of the KS states of the high-spin reference state \cite{myneni2017calculation}. (More sophisticated DFT approximations have also been developed \cite{wang1995evaluation,Cohen2007}.) %
In Fig. \ref{fig:ethylene_ref_s2}, we see that $\langle \hat{S}^2 \rangle_0$ deviates by at most 0.1\% from 2.0 for ethylene under any torsion angle, as expected for the triplet spin symmetry of the high-spin reference state. Likewise, $\langle \hat{S}^2 \rangle_0$ for the triplet ground state of the NV$^-$ center is 2.05. These results are consistent with a DFT study \cite{tada2021estimation} finding that solids generally have more spin contamination than molecules.

 For spin-flip excitations from the high-spin reference state, we compute the difference ``$\Delta \langle \hat{S}^2\rangle$''between $\langle \hat{S}^2 \rangle$ for the $I$'th spin-flipped excitation and the reference state: $\Delta \langle \hat{S}^2\rangle = \langle \hat{S}^2 \rangle_I - \langle \hat{S}^2 \rangle_0$.  (It is standard to subscript $\Delta \langle \hat{S}^2\rangle$ with the label for the state $I$ for which it is computed; we will be suppressing this subscript throughout.) While the formula for $\Delta \langle\hat{S}^2 \rangle$ in the context of SF-TDDFT was previously published as Eq. A10 in \cite{li2011spin}, was evidently implemented in codes for use in calculations such as \cite{xu2014testing}, and was used for SF-BSE in \cite{monino2021spin}, the derivation does not seem to have been shown in detail and so we work through the derivation here.
 
The eigenvectors of the Bethe-Salpeter Equation are the electron-hole amplitudes \cite{Strinati} that are required to construct the two-particle reduced difference density matrix $\Delta \Gamma$ \cite{ipatov2009excited}:
\begin{eqnarray}
& \Gamma^0_{pq,rs} & = \langle N,0| r^{\dagger}s^{\dagger}qp| N,0 \rangle \\
& \Gamma^I_{pq,rs} & = \langle N,I | r^{\dagger}s^{\dagger}qp |N,I\rangle \\
& \Delta \Gamma_{pq,rs} & = \Gamma^I_{pq,rs} - \Gamma^0_{pq,rs},
\end{eqnarray}
where $p$, $q$, $r$, and $s$ are the annihilation operators for one-electron states $\phi_p$, $\phi_q$, $\phi_r$, and $\phi_s$, respectively; $|N,0\rangle$ is the many-body high-spin reference state; and $|N,I\rangle$ is the $I$'th many-body excited state.

We use the formalism developed in \cite{ipatov2009excited}, originally for spin-conserving transitions and obtained from considering time-dependent Hartree-Fock, and apply it to spin-flipping transitions. Within this formalism, we evaluate the $I$'th excited-state's $\langle \hat{S}^2 \rangle_I$ indirectly by adding $\langle \hat{S}^2 \rangle_0$ 
to  $\Delta \langle \hat{S}^2 \rangle_I$, which is computed using from the ``super-operator'' approach \cite{ipatov2009excited,myneni2017calculation,li2011spin}:
\begin{eqnarray}
\label{eq:delta_gamma}
&   \Delta \langle \hat{S}^2 \rangle =  1 - 2M^{\textrm{H.S. Ref}}_S + \Delta \langle \hat{\mathcal{P}}_{\uparrow\downarrow}\rangle \, ,& \\
& \Delta \langle \hat{\mathcal{P}}_{\uparrow\downarrow}\rangle = \sum_{r,s,\bar{p},\bar{q}} \, \Delta \Gamma_{r\uparrow,\bar{q}\downarrow,\bar{p}\downarrow,s\uparrow}\langle s | \bar{q} \rangle \langle \bar{p} | r \rangle \,  ,& \\
&  \Delta \Gamma_{r\uparrow,\bar{q}\downarrow,\bar{p}\downarrow,s\uparrow}  = \sum_{i\uparrow,\bar{a}\downarrow,j\uparrow,\bar{b}\downarrow} \left(A^I_{i\uparrow,\bar{a}\downarrow} \right)^{*}  A^I_{j\uparrow,\bar{b}\downarrow} \times &\\ \nonumber 
& \,\,\,\,\, \,\,\,\, \langle N, 0 | \left[ i^{\dagger}_{\uparrow} \bar{a}_{\downarrow} \, , \, \left[ \bar{p}^{\dagger}_{\downarrow} s^{\dagger}_{\uparrow} \bar{q}_{\downarrow} r_{\uparrow} \, , \, \bar{b}^{\dagger}_{\downarrow} j_{\uparrow} \right] \right] | N, 0 \rangle \, . &  \nonumber
\end{eqnarray}
The constant term $1 - 2M^{\textrm{H.S. Ref}}_S$ appears due to the change in the number of up- and down-spin electrons in the spin-flipped excited state: 
\begin{eqnarray}
& 1 - 2M^{\textrm{H.S. Ref}}_S  & \\ \nonumber
& \,\,\, = \left( M^I_S(M^I_S+1) + N^I_{\downarrow} \right) - \left(M^{\textrm{H.S. Ref}}(M^{\textrm{H.S. Ref}}+1) + N^{\textrm{H.S. Ref}}_{\downarrow} \right) \, , & \\
& M^I_S = M^{\textrm{H.S. Ref}} - 1 \, & \\
& N^I_{\downarrow} = N^{\textrm{H.S. Ref}} + 1 & \, .
\end{eqnarray}
We use the convention from the quantum chemistry literature that $i$, $j$, $k$ are indices for creation/annihilation operators for occupied orbitals or bands; $a$, $b$, unoccupied orbitals or bands; and $p$, $q$, $r$, and $s$, either occupied or unoccupied. Again, the over-bar indicates the possibility of the use of unrestricted orbitals or, equivalently, spin-polarized bands.

\begin{figure}
\centering
\includegraphics[scale=0.5]{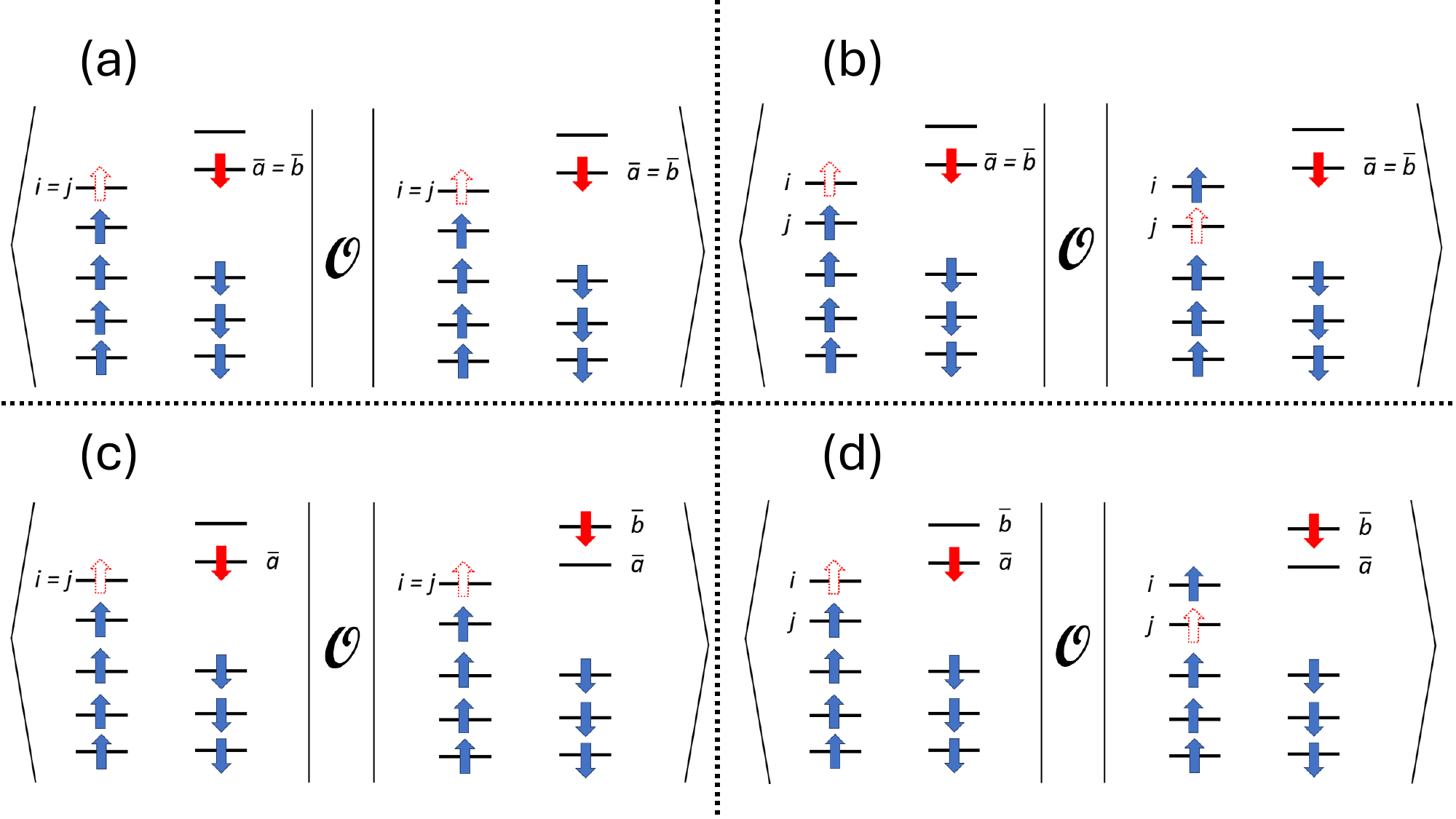}
\caption{The four cases (as in Eq. \ref{eq:partition}) to consider for calculating matrix elements of the nested commutators of creation/annihilation operators, labelled as $\mathcal{O}$ in the figure, necessary for the calculation of $\Delta \langle \hat{S}^2 \rangle$. The indices $i$ and $j$ represent the newly unoccupied state upon the spin-flipping excitation, for the bra and ket, respectively, and $\bar{a}$ and $\bar{b}$ represent the newly occupied state for the bra and ket, respectively. (a) The single-particle transitions for the bra and ket involve the same up-spin orbital (``$i = j$'' ) and same down-spin orbital (``$\bar{a} = \bar{b}$''), (b) different up-spin orbitals but the same down-spin orbitals (``$i \neq j$'' but ``$\bar{a} = \bar{b}$'' ), (c) the same up-spin orbitals but different down-spin orbitals (``$i = j$'' but ``$\bar{a} \neq \bar{b}$'' ), and (d) different up-spin and down-spin orbitals (``$i \neq j$'' and ``$\bar{a} \neq \bar{b}$'').
}
\label{fig:Delta_S2_sketch}
\end{figure}

The nested commutators were confirmed to be evaluated correctly with the assistance of the SNEG \cite{SNEG} software, which can perform symbolic evaluation of second-quantization-operator expressions. The first general result for the matrix element is
\begin{eqnarray}
\label{eq:expanded_commutator}
&    \langle N, 0 | \left[ i^{\dagger}_{\uparrow} \bar{a}_{\downarrow} \, , \, \left[ \bar{p}^{\dagger}_{\downarrow} s^{\dagger}_{\uparrow} \bar{q}_{\downarrow} r_{\uparrow} \, , \, \bar{b}^{\dagger}_{\downarrow} j_{\uparrow} \right] \right] | N, 0 \rangle & \\ \nonumber
& =  \langle N, 0 | i^{\dagger}_{\uparrow} \bar{a}_{\downarrow} \bar{p}^{\dagger}_{\downarrow} s^{\dagger}_{\uparrow} \bar{q}_{\downarrow} r_{\uparrow} \bar{b}^{\dagger}_{\downarrow} j_{\uparrow}  | N, 0 \rangle - \langle N, 0 | i^{\dagger}_{\uparrow} \bar{a}_{\downarrow} \bar{b}^{\dagger}_{\downarrow} j_{\uparrow} \bar{p}^{\dagger}_{\downarrow} s^{\dagger}_{\uparrow} \bar{q}_{\downarrow} r_{\uparrow} | N, 0 \rangle& \, ,
\end{eqnarray}
after recognizing that terms with $\bar{a}_{\downarrow} |N,0\rangle = 0$. %
As in \cite{wang1995evaluation}, we explicitly consider four cases (see Fig. \ref{fig:Delta_S2_sketch}): (1) $i = j$, $\bar{a} = \bar{b}$, (2a) $i \neq j$, $\bar{a} = \bar{b}$, (2b) $i = j$, $\bar{a} \neq \bar{b}$, and (3) $i \neq j$, $\bar{a} \neq \bar{b}$, thus partitioning the expression as follows:
\begin{eqnarray}
\label{eq:partition}
\Delta \langle \hat{S}^2 \rangle = & 1 - 2M^{\textrm{H.S. Ref}}_S + \sum_{i,\bar{a},j,\bar{b}} \left(A^I_{i\uparrow,\bar{a}\downarrow}\right)^{*} A^I_{j\uparrow,\bar{b}\downarrow} \bigg. \times & \\ \nonumber
& \bigg\{ \delta_{i,j}\delta_{\bar{a},\bar{b}}  \langle \mathcal{O}\rangle^{i=j,\bar{a}=\bar{b}} + (1-\delta_{i,j})\delta_{\bar{a},\bar{b}} \langle \mathcal{O}\rangle^{i \neq j,\bar{a}=\bar{b}} & \\ \nonumber
& \bigg. + \delta_{i,j}(1 - \delta_{\bar{a},\bar{b}})  \langle \mathcal{O}\rangle^{i = j,\bar{a} \neq \bar{b}} + (1 - \delta_{i,j})(1 - \delta_{\bar{a},\bar{b}})  \langle \mathcal{O}\rangle^{i \neq j,\bar{a} \neq \bar{b}} \bigg\} \, ,&
\end{eqnarray}
where 
\begin{equation}
\label{eq:misc_operator}
\langle \mathcal{O} \rangle = \sum_{r,s,\bar{p},\bar{q}}\langle N, 0 | \left[ i^{\dagger}_{\uparrow} \bar{a}_{\downarrow} \, , \, \left[ \bar{p}^{\dagger}_{\downarrow} s^{\dagger}_{\uparrow} \bar{q}_{\downarrow} r_{\uparrow} \, , \, \bar{b}^{\dagger}_{\downarrow} j_{\uparrow} \right] \right] | N, 0 \rangle \langle s | \bar{q} \rangle \langle \bar{p} | r \rangle \, .
\end{equation}

We consider explicitly the matrix element of the nested commutators in the contribution to 
$\Delta \langle \hat{S}^2 \rangle$
for ``Case 1,'' when $i=j$ and $\bar{a}=\bar{b}$:
\begin{eqnarray}
& \langle N, 0 | i^{\dagger}_{\uparrow} \bar{a}_{\downarrow} \bar{p}^{\dagger}_{\downarrow} s^{\dagger}_{\uparrow} \bar{q}_{\downarrow} r_{\uparrow} \bar{b}^{\dagger}_{\downarrow} j_{\uparrow}  | N, 0 \rangle - \langle N, 0 | i^{\dagger}_{\uparrow} \bar{a}_{\downarrow} \bar{b}^{\dagger}_{\downarrow} j_{\uparrow} \bar{p}^{\dagger}_{\downarrow} s^{\dagger}_{\uparrow} \bar{q}_{\downarrow} r_{\uparrow} | N, 0 \rangle \\ \nonumber
& = \langle N, 0 | i^{\dagger}_{\uparrow} \bar{a}_{\downarrow} \bar{p}^{\dagger}_{\downarrow} s^{\dagger}_{\uparrow} \bar{q}_{\downarrow} r_{\uparrow} \bar{a}^{\dagger}_{\downarrow} i_{\uparrow}  | N, 0 \rangle - \langle N, 0 | i^{\dagger}_{\uparrow} \bar{a}_{\downarrow} \bar{a}^{\dagger}_{\downarrow} i_{\uparrow} \bar{p}^{\dagger}_{\downarrow} s^{\dagger}_{\uparrow} \bar{q}_{\downarrow} r_{\uparrow} | N, 0 \rangle & \\ \nonumber
& = -\langle N, 0 | i^{\dagger}_{\uparrow} \left(\delta(\bar{p},\bar{a}) - \bar{p}^{\dagger}_{\downarrow} \bar{a}_{\downarrow}\right) s^{\dagger}_{\uparrow} \left(\delta(\bar{q},\bar{a}) - \bar{a}^{\dagger}_{\downarrow}\bar{q}_{\downarrow}\right)  r_{\uparrow} i_{\uparrow}  | N, 0 \rangle & \\ \nonumber
& \,\, + \langle N, 0 | i^{\dagger}_{\uparrow} \left( \delta(\bar{a},\bar{a}) - \bar{a}^{\dagger}_{\downarrow}\bar{a}_{\downarrow} \right)  \bar{p}^{\dagger}_{\downarrow} \left( \delta(s,i) - s^{\dagger}_{\uparrow} i_{\uparrow}  \right) \bar{q}_{\downarrow} r_{\uparrow} | N, 0 \rangle  & \\ \nonumber
& = -\langle N, 0 | \delta(\bar{p},\bar{a}) \delta(\bar{q},\bar{a})i^{\dagger}_{\uparrow}  s^{\dagger}_{\uparrow}   r_{\uparrow} i_{\uparrow}  | N, 0 \rangle + \langle N, 0 | \delta(\bar{p},\bar{a}) i^{\dagger}_{\uparrow}  s^{\dagger}_{\uparrow} \bar{a}^{\dagger}_{\downarrow} \bar{q}_{\downarrow}  r_{\uparrow} i_{\uparrow}  | N, 0 \rangle & \\ \nonumber
& \,\, - \langle N, 0 | \delta(\bar{q},\bar{a}) i^{\dagger}_{\uparrow}   \bar{p}^{\dagger}_{\downarrow}s^{\dagger}_{\uparrow} \bar{a}_{\downarrow}  r_{\uparrow} i_{\uparrow}  | N, 0 \rangle + \langle N, 0 |  i^{\dagger}_{\uparrow}   \bar{p}^{\dagger}_{\downarrow}s^{\dagger}_{\uparrow} \left(\delta(\bar{a},\bar{a}) -  \bar{a}^{\dagger}_{\downarrow}\bar{a}_{\downarrow} \right) \bar{q}_{\downarrow} r_{\uparrow} i_{\uparrow}  | N, 0 \rangle & \\ \nonumber
& \,\, - \langle N, 0 | \delta(\bar{a},\bar{a}) i^{\dagger}_{\uparrow}     \bar{p}^{\dagger}_{\downarrow} s^{\dagger}_{\uparrow}  \bar{q}_{\downarrow} r_{\uparrow} i_{\uparrow} | N, 0 \rangle + \langle N, 0 | \delta(\bar{a},\bar{a}) \delta(s,i) i^{\dagger}_{\uparrow}     \bar{p}^{\dagger}_{\downarrow}  \bar{q}_{\downarrow} r_{\uparrow}  | N, 0 \rangle \nonumber .
\end{eqnarray}
Again, terms with $a_{\downarrow}|N,0\rangle = 0$ (and likewise with its complex conjugate), which leaves us with just four terms:
\begin{eqnarray}
\label{eq:case1_intermediate}
& = \langle N, 0 | \delta(\bar{a},\bar{a}) i^{\dagger}_{\uparrow} \bar{p}^{\dagger}_{\downarrow} s^{\dagger}_{\uparrow} \bar{q}_{\downarrow} r_{\uparrow} i_{\uparrow}  | N, 0 \rangle - \langle N, 0 | \delta(\bar{a},\bar{p})\delta(\bar{a},\bar{q}) i^{\dagger}_{\uparrow} s^{\dagger}_{\uparrow}  r_{\uparrow} i_{\uparrow}  | N, 0 \rangle & \\ \nonumber
& \,\,- \langle N, 0 | \delta(\bar{a},\bar{a}) i^{\dagger}_{\uparrow} \bar{p}^{\dagger}_{\downarrow} s^{\dagger}_{\uparrow} \bar{q}_{\downarrow} r_{\uparrow} i_{\uparrow}  | N, 0 \rangle + \langle N, 0 | \delta(\bar{a},\bar{a})\delta(i,s) i^{\dagger}_{\uparrow}  \bar{p}^{\dagger}_{\downarrow} \bar{q}_{\downarrow} r_{\uparrow} | N, 0 \rangle & \nonumber.
\end{eqnarray}
The first and third terms cancel, giving
\begin{equation}
\delta(\bar{a},\bar{a})\delta(s,i)\delta(r,i)\delta(\bar{p},\bar{q}) - \delta(\bar{p},\bar{a})\delta(\bar{q},\bar{a})\delta(i,i)\delta(s,r)(1-\delta(r,i))\, .
\end{equation}

Placing these Kroenecker deltas into Eq. \ref{eq:misc_operator}, $ \langle \mathcal{O} \rangle$ for this first case gives
\begin{eqnarray}
&  \langle  \mathcal{O}\rangle ^{i=j,\bar{a}=\bar{b}} & \\ \nonumber
&  \,\,\,\,\, =  \sum_{r,s,\bar{p},\bar{q}} \, \left( \,\delta(\bar{a},\bar{a})\delta(s,i)\delta(r,i)\delta(\bar{p},\bar{q})  - \delta(\bar{p},\bar{a})\delta(\bar{q},\bar{a})\delta(i,i)\delta(s,r)(1-\delta(r,i) \right) \langle s | \bar{q}\rangle \langle \bar{p} | r \rangle\, \, , & \\ \nonumber
&  \,\,\,\,\,  =   \sum_{\bar{p}} \, 
 \langle i | \bar{p}\rangle \langle \bar{p} | i \rangle  -  \sum_{r \neq i} \, 
 \langle r | \bar{a}\rangle \langle \bar{a} | r \rangle\, \, , & \\ \nonumber
& \,\,\,\,\,  = -\left( \sum_k |\langle k |\bar{a} \rangle |^2 - \sum_{\bar{k}} |\langle i |\bar{k} \rangle |^2 - |\langle i |\bar{a} \rangle |^2  \right) \, , & \nonumber
\end{eqnarray}
with the convention that $k$ and $\bar{k}$ are the occupied states from the high-spin reference state, for either spin channel. (With this definition for $k$, the third term enforces the condition from the second term in Eq. \ref{eq:case1_intermediate} that the states labeled $r$  are prohibited from including $i$.)

We now consider the derivation for ``Case 2a,'' when $i \neq j$ and $\bar{a}=\bar{b}$:
\begin{eqnarray}
& \langle N, 0 | i^{\dagger}_{\uparrow} \bar{a}_{\downarrow} \bar{p}^{\dagger}_{\downarrow} s^{\dagger}_{\uparrow} \bar{q}_{\downarrow} r_{\uparrow} \bar{a}^{\dagger}_{\downarrow} j_{\uparrow}  | N, 0 \rangle - \langle N, 0 | i^{\dagger}_{\uparrow} \bar{a}_{\downarrow} \bar{a}^{\dagger}_{\downarrow} j_{\uparrow} \bar{p}^{\dagger}_{\downarrow} s^{\dagger}_{\uparrow} \bar{q}_{\downarrow} r_{\uparrow} | N, 0 \rangle & \\ \nonumber
& = \langle N, 0 | \delta(\bar{a},\bar{a}) i^{\dagger}_{\uparrow} \bar{p}^{\dagger}_{\downarrow} s^{\dagger}_{\uparrow} \bar{q}_{\downarrow} r_{\uparrow} j_{\uparrow}  | N, 0 \rangle - \langle N, 0 | \delta(\bar{a},\bar{p})\delta(\bar{a},\bar{q}) i^{\dagger}_{\uparrow} s^{\dagger}_{\uparrow}  r_{\uparrow} j_{\uparrow}  | N, 0 \rangle & \\ \nonumber
& - \langle N, 0 | \delta(\bar{a},\bar{a}) i^{\dagger}_{\uparrow} \bar{p}^{\dagger}_{\downarrow} s^{\dagger}_{\uparrow} \bar{q}_{\downarrow} r_{\uparrow} j_{\uparrow}  | N, 0 \rangle + \langle N, 0 | \delta(\bar{a},\bar{a})\delta(j,s) i^{\dagger}_{\uparrow}  \bar{p}^{\dagger}_{\downarrow} \bar{q}_{\downarrow} r_{\uparrow} | N, 0 \rangle & \nonumber .
\end{eqnarray}
The first and third terms cancel, and the remaining terms are
\begin{eqnarray}
&  - \langle N, 0 | \delta(\bar{a},\bar{p})\delta(\bar{a},\bar{q}) i^{\dagger}_{\uparrow} s^{\dagger}_{\uparrow}  r_{\uparrow} j_{\uparrow}  | N, 0 \rangle  + \langle N, 0 | \delta(\bar{a},\bar{a})\delta(j,s) i^{\dagger}_{\uparrow}  \bar{p}^{\dagger}_{\downarrow} \bar{q}_{\downarrow} r_{\uparrow} | N, 0 \rangle & \\ \nonumber
& = \delta(\bar{a},\bar{a}) \delta(j,s)\delta(i,r) \delta(\bar{p},\bar{q}) - \delta(\bar{a},\bar{p})\delta(\bar{a},\bar{q})\delta(i,j)\delta(s,r) & \,. \nonumber
\end{eqnarray}
However, $i \neq j$, so we only have one non-zero contribution for the case $i \neq j$ and $\bar{a} = \bar{b}$: 
\begin{equation} \delta(\bar{a},\bar{a})\delta(j,s)\delta(i,r)\delta(\bar{p},\bar{q}).
\end{equation}
The ``Case 2a'' contribution to $\Delta \langle \hat{S}^2 \rangle$ is found, then, to be 
\begin{eqnarray}
\langle \mathcal{O} \rangle^{i \neq j,\bar{a}=\bar{b}} & =\, \sum_{r,s,\bar{p},\bar{q}} \, \delta(\bar{a},\bar{a})\delta(j,s)\delta(i,r)\delta(\bar{p},\bar{q}) \langle s | \bar{q}\rangle \langle \bar{p} | r \rangle\, \, , & \\ \nonumber
& = \, \sum_{\bar{k}} \,  \langle j | \bar{k}\rangle \langle \bar{k} | i \rangle\, \, . &
\end{eqnarray}

Similarly, we consider the derivation for ``Case 2b,'' when $i = j$ and $\bar{a} \neq \bar{b}$:
\begin{equation}
\langle N, 0 | i^{\dagger}_{\uparrow} \bar{a}_{\downarrow} \bar{p}^{\dagger}_{\downarrow} s^{\dagger}_{\uparrow} \bar{q}_{\downarrow} r_{\uparrow} \bar{b}^{\dagger}_{\downarrow} i_{\uparrow}  | N, 0 \rangle - \langle N, 0 | i^{\dagger}_{\uparrow} \bar{a}_{\downarrow} \bar{b}^{\dagger}_{\downarrow} i_{\uparrow} \bar{p}^{\dagger}_{\downarrow} s^{\dagger}_{\uparrow} \bar{q}_{\downarrow} r_{\uparrow} | N, 0 \rangle \, .
\end{equation}
We remark immediately that the second term is necessarily zero due to the position of the $\bar{b}^{\dagger}_{\downarrow}$ operator. Its commutation with $\bar{a}_{\downarrow}$ gives the Kroenecker delta $\delta(\bar{a},\bar{b})$ which is zero definition of Case 2b, and the null $\left( \bar{b}_{\downarrow}|N,0\rangle\right)^{\dagger}$.
The non-zero portion may be reduced:
\begin{eqnarray}
& = -\langle N, 0 | i^{\dagger}_{\uparrow} \left( \delta(\bar{p},\bar{a}) -  \bar{p}^{\dagger}_{\downarrow}\bar{a}_{\downarrow}\right) s^{\dagger}_{\uparrow} \bar{q}_{\downarrow} \bar{b}^{\dagger}_{\downarrow} r_{\uparrow} i_{\uparrow}  | N, 0 \rangle &\\ \nonumber
& = -\langle N, 0 | i^{\dagger}_{\uparrow} \left( \delta(\bar{p},\bar{a}) -  \bar{p}^{\dagger}_{\downarrow}\bar{a}_{\downarrow}\right) s^{\dagger}_{\uparrow} \left(\delta(\bar{b},\bar{q}) -  \bar{b}^{\dagger}_{\downarrow}\bar{q}_{\downarrow} \right) r_{\uparrow} i_{\uparrow}  | N, 0 \rangle & \\ \nonumber
& = -\langle N, 0 | \delta(\bar{p},\bar{a})\delta(\bar{q},\bar{b})i^{\dagger}_{\uparrow}  s^{\dagger}_{\uparrow}  r_{\uparrow} i_{\uparrow}  | N, 0 \rangle & \\ \nonumber
& \,\, + \langle N, 0 | \delta(\bar{p},\bar{a}) i^{\dagger}_{\uparrow} s^{\dagger}_{\uparrow} \bar{b}^{\dagger}_{\downarrow}\bar{q}_{\downarrow} r_{\uparrow} i_{\uparrow}  | N, 0 \rangle & \\ \nonumber
& \,\, -\langle N, 0 | \delta(\bar{b},\bar{q}) i^{\dagger}_{\uparrow}\bar{p}^{\dagger}_{\downarrow}s^{\dagger}_{\uparrow}\bar{a}_{\downarrow}  r_{\uparrow} i_{\uparrow}  | N, 0 \rangle & \\ \nonumber
& \,\, + \langle N, 0 | i^{\dagger}_{\uparrow}  \bar{p}^{\dagger}_{\downarrow}s^{\dagger}_{\uparrow}\bar{a}_{\downarrow}   \bar{b}^{\dagger}_{\downarrow}\bar{q}_{\downarrow} r_{\uparrow} i_{\uparrow}  | N, 0 \rangle  \,. & \nonumber
\end{eqnarray}
Of these final four terms, only the first is non-zero. The remaining three are zero due to the conditions $\bar{a}_{\downarrow}|N,0\rangle = 0$, $\left(\bar{b}_{\downarrow}|N,0\rangle\right)^{\dagger} = 0$, and $\delta(\bar{a},\bar{b}) = 0$ for $\bar{a} \neq \bar{b}$. Thus ``Case 2b,'' when $i = j$ and $\bar{a} \neq \bar{b}$, gives as the matrix element
\begin{equation}
-\delta(\bar{p},\bar{a}) \delta(\bar{q},\bar{b}) \delta(s,r) \delta(i,i)(1-\delta(r,i)) .
\end{equation} 
Using this, we find the ``Case 2b'' contribution to $\Delta \langle \hat{S}^2 \rangle$: 
\begin{eqnarray}
\langle \mathcal{O} \rangle^{i = j,\bar{a} \neq \bar{b}} & = -\sum_{r\neq i,s,\bar{p},\bar{q}} \, \delta(\bar{p},\bar{a}) \delta(\bar{q},\bar{b}) \delta(s,r) \delta(i,i)\langle s | \bar{q}\rangle \langle \bar{p} | r \rangle\, \, & \\ \nonumber
&  = - \sum_{k} \,  \langle k | \bar{b}\rangle \langle \bar{a} | k \rangle\, - \langle i | \bar{b} \rangle \langle \bar{a} | i \rangle\, , &
\end{eqnarray}
where, again, the contribution from $i$ in the sum over $k$ must be subtracted out for the same reason as in ``Case 1.''

Finally, we derive the contribution for ``Case 3,'' when $ i \neq j$ and $\bar{a} \neq \bar{b}$:
\begin{equation}
\langle N, 0 | i^{\dagger}_{\uparrow} \bar{a}_{\downarrow} \bar{p}^{\dagger}_{\downarrow} s^{\dagger}_{\uparrow} \bar{q}_{\downarrow} r_{\uparrow} \bar{b}^{\dagger}_{\downarrow} j_{\uparrow}  | N, 0 \rangle - \langle N, 0 | i^{\dagger}_{\uparrow} \bar{a}_{\downarrow} \bar{b}^{\dagger}_{\downarrow} j_{\uparrow} \bar{p}^{\dagger}_{\downarrow} s^{\dagger}_{\uparrow} \bar{q}_{\downarrow} r_{\uparrow} | N, 0 \rangle \, .
\end{equation}
As in the previous case, the second term is necessarily zero. We then have
\begin{eqnarray}
& = -\langle N, 0 | i^{\dagger}_{\uparrow} \left( \delta(\bar{p},\bar{a}) -  \bar{p}^{\dagger}_{\downarrow}\bar{a}_{\downarrow}\right) s^{\dagger}_{\uparrow} \left(\delta(\bar{b},\bar{q}) -  \bar{b}^{\dagger}_{\downarrow}\bar{q}_{\downarrow} \right) r_{\uparrow} j_{\uparrow}  | N, 0 \rangle & \\ \nonumber
&  = -\langle N, 0 |\delta(\bar{p},\bar{a}) \delta(\bar{b},\bar{q})i^{\dagger}_{\uparrow}  s^{\dagger}_{\uparrow}  r_{\uparrow} j_{\uparrow}  | N, 0 \rangle & \\ \nonumber
& = \langle N, 0 |\delta(\bar{p},\bar{a}) \delta(\bar{b},\bar{q})s^{\dagger}_{\uparrow} i^{\dagger}_{\uparrow}    r_{\uparrow} j_{\uparrow}  | N, 0 \rangle & \nonumber
\end{eqnarray}
with the other terms necessarily zero, for the same reasons as in the previous case.
The final case, then, gives us for the matrix element
\begin{equation}
\delta(\bar{p},\bar{a}) \delta(\bar{q},\bar{b}) \delta(r,i) \delta(s,j).
\end{equation}
With this, the ``Case 3'' contribution to $\Delta \langle \hat{S}^2 \rangle$ is 
\begin{eqnarray}
\langle \mathcal{O}\rangle^{i \neq j,\bar{a} \neq \bar{b}\textrm{ only}} & = \sum_{r,s,\bar{p},\bar{q}} \, \delta(\bar{p},\bar{a}) \delta(\bar{q},\bar{b}) \delta(r,i) \delta(s,j)\langle s | \bar{q}\rangle \langle \bar{p} | r \rangle\, \, & \\ \nonumber
& = \langle j | \bar{b}\rangle \langle \bar{a} | i \rangle\, \, . &
\end{eqnarray}

Collecting all of the contributions from the different cases as per Eq. \ref{eq:partition}, we arrive at the complete expression for $\Delta \langle \hat{S}^2 \rangle$:
\begin{eqnarray}
\label{eq:delta_s2}
\Delta \langle \hat{S}^2 \rangle = & 1 - 2M^{\textrm{H.S. Ref}}_S + \sum_{i,\bar{a},j,\bar{b}} \left(A^I_{i\uparrow,\bar{a}\downarrow}\right)^{*} A^I_{j\uparrow,\bar{b}\downarrow} \bigg. \times & \\ \nonumber
&  \bigg\{  -\delta_{i,j}\delta_{\bar{a},\bar{b}} \left( \sum_k |\langle k |\bar{a} \rangle |^2 - \sum_{\bar{k}} |\langle i |\bar{k} \rangle |^2 - |\langle i |\bar{a} \rangle |^2  \right) &\\ \nonumber
& + (1-\delta_{i,j})\delta_{\bar{a},\bar{b}} \sum_{\bar{k}} \langle j | \bar{k} \rangle \langle \bar{k} | i \rangle - \delta_{i,j}(1 - \delta_{\bar{a},\bar{b}})\left(\sum_k  \langle k | \bar{b} \rangle \langle \bar{a} | k \rangle - \langle i | \bar{b} \rangle \langle \bar{a} | i \rangle \right)& \\ \nonumber
& \bigg. + (1 - \delta_{i,j})(1 - \delta_{\bar{a},\bar{b}}) \langle j | \bar{b} \rangle \langle \bar{a} | i \rangle \bigg\}  \, . &
\end{eqnarray}
This equation has an important property of gauge-invariance which can be explicitly verified: a phase change of an individual orbital does not affect the final result, provided that the coefficients $A$ have the counteracting change that keeps the quasiparticle wavefunction $|\Psi^{\textit{I}}\rangle$ constant.

\section{Example application: The defect states of the NV$^{-}$ center}
\label{sec:example}

The general equation for $\Delta \langle \hat{S}^2 \rangle$ is clarified by considering a simple and illustrative example. We will consider a by-hand calculation of Eq. \ref{eq:delta_s2} for the $S=1, M_S = 0$ ground state of the NV$^{-}$ center, with a minimal basis set, and with the same orbitals for the different spin channels (as in a ``spin-restricted'' calculation).

\begin{figure}
\centering
\includegraphics[scale=0.30]{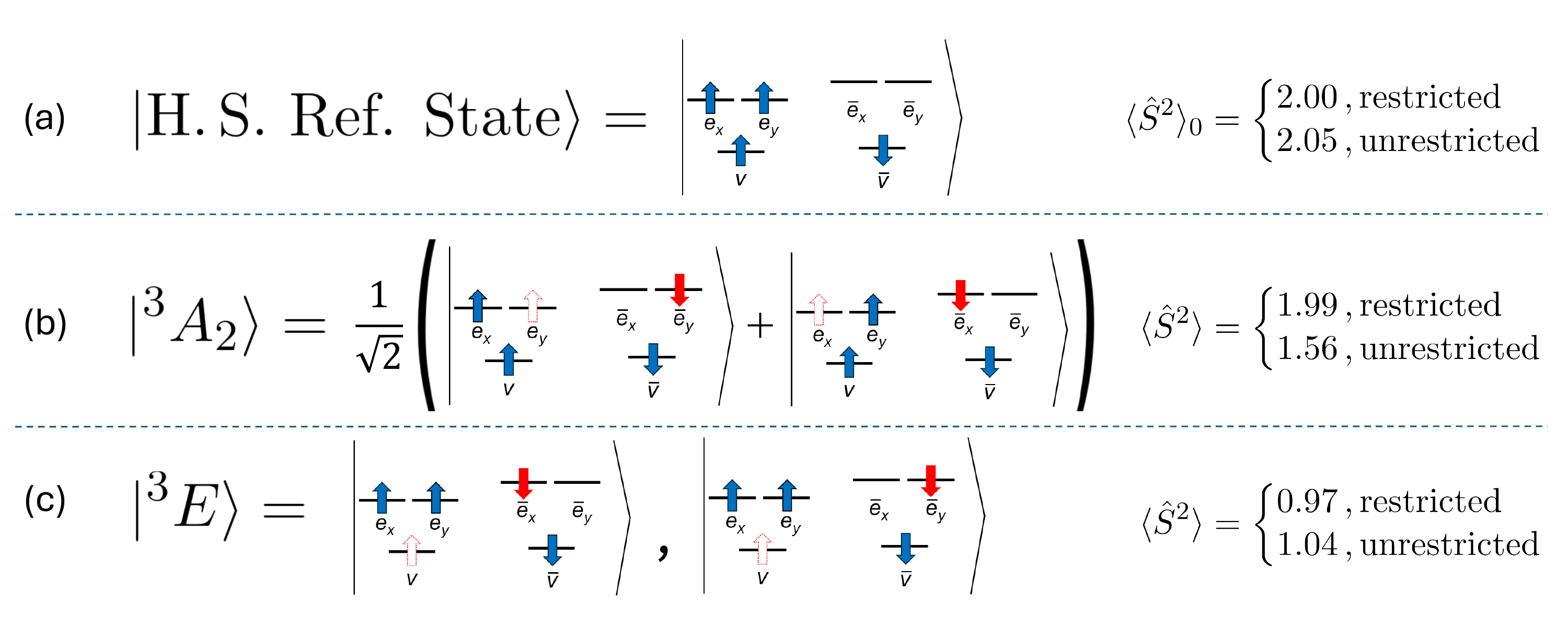}
\caption{The electron configurations and values of $\langle \hat{S}^2 \rangle$ as calculated in SF-BSE for (a) the high-spin reference state, (b) the $^3A_2$ (with $M_S = 0$) ground state, and (c) the $^3E$ excited state for the NV$^{-}$ center in diamond. Values of $\langle \hat{S}^2 \rangle$ are presented for both the spin-restricted and spin-unrestricted cases.}
\label{fig:nvminus_s2_states}
\end{figure}

In a more complete description, the NV$^{-}$ center has four electrons available to occupy six in-gap orbitals \cite{maze2011properties}: $|v\uparrow\rangle$, $|\bar{v}\downarrow\rangle$ , $|e_x\uparrow\rangle$, $|\bar{e}_x\downarrow\rangle$ , $|e_y\uparrow\rangle$, $|\bar{e}_y\downarrow\rangle$, where $e_x$ and $e_y$ are degenerate single-particle states. The high-spin (triplet) reference state $| v\uparrow,\bar{v}\downarrow, e_x\uparrow, e_y\uparrow\rangle$ has $S=1, M_S =1$. From this high-spin reference state, we find from actual SF-BSE calculations the $S=1, M_S = 0$ triplet state

\begin{equation}
|^3A_2\rangle = 0.70 | e_x \bar{e}_y\rangle + 0.70 | e_y \bar{e}_x\rangle \, + \, ... \, ,
\end{equation}
omitting the $v$ states as well as the spins (the overbar denoting the down-spin orbital, and the lack of an overbar denoting the up-spin).
While the small deviation of the coefficients from $1/\sqrt{2}$ in the presented eigenvector for $|^3A_2\rangle$ is from minor contributions of other excitations, these are found to be less than 0.1; we therefore round the coefficients to $1/\sqrt{2}$ for an analytic calculation of $\langle \hat{S}^2 \rangle$ for this triplet state. See Fig. \ref{fig:nvminus_s2_states} for an illustration of the high-spin reference state, $|^3A_2\rangle$, and the degenerate excited triplet states with maximal spin contamination, $|^3E\rangle$.

To first calculate $\langle \hat{S}^2 \rangle_0$ for the high-spin reference state, we reiterate our assumption in our simple model that we have the same orbitals for the different spin channels, and we apply the L{\"o}wdin formula (Eq. \ref{eq:lowdin}):
\begin{eqnarray}
\langle \hat{S}^2 \rangle_0 & = \left( \frac{2}{2} \right) \left( \frac{2}{2} + 1 \right) + 1 - 1 =2 
\end{eqnarray}
since only the $v$ orbitals have both spin channels occupied. The use of spin-unrestricted orbitals causes only small deviations from the expected value, 2, with the calculated $\langle \hat{S}^2 \rangle \approx 2.05$.

We illustrate the calculation of $\Delta \langle \hat{S}^2 \rangle$ for the $|^3A_2\rangle $ state (with $S = 1, M_S = 0$), using the simplified spin-restricted orbitals, with the exact coefficients of $1/\sqrt{2}$. This state, then, is
\begin{equation}
|^3A_2\rangle = \frac{1}{\sqrt{2}}| e_x \bar{e}_y\rangle + \frac{1}{\sqrt{2}} | e_y \bar{e}_x\rangle \,. 
\end{equation}
The calculation of $\Delta \langle \hat{S}^2 \rangle$ can most readily be thought of in terms of the ``four cases'' for $\Delta \Gamma$ in Eq. \ref{eq:delta_gamma}.

In this minimal basis set (where we even ignore the ``$v$'' in-gap orbitals), the bra $\langle ^3A_2 |$ and the ket $| ^3A_2 \rangle$ each are composed of two spin-flip transitions (with respect to the high-spin reference state $|e_x,e_y\rangle$), and each of these transitions has an amplitude $A_{i,\bar{a}} = \frac{1}{\sqrt{2}}$:
\begin{equation}
| ^3A_2 \rangle = \frac{1}{\sqrt{2}} \bar{e}^{\dagger}_{x} e_{x} |e_x,e_y\rangle + \frac{1}{\sqrt{2}} \bar{e}^{\dagger}_{y} e_{y} |e_x,e_y \rangle \nonumber ,
\end{equation}
where $e_x$ is the annihilation operator for the $e_x$ band, etc.
Similarly, the bra state can be written as
\begin{equation}
\langle ^3A_2 | = \frac{1}{\sqrt{2}} \langle e_x,e_y| \left(\bar{e}^{\dagger}_{x} e_{x}\right)^{\dagger}  + \frac{1}{\sqrt{2}} \langle e_x,e_y| \left(\bar{e}^{\dagger}_y e_y\right)^{\dagger}.  \nonumber 
\end{equation}

In calculating Eq. \ref{eq:delta_s2}, we see that we have only have two of the possible four cases: $i = j$ and $\bar{a} = \bar{b}$, and $i \neq j$ and $\bar{a} \neq \bar{b}$. Let us apply the cases individually.
When $i = j = e_x$ and $\bar{a} = \bar{b} = \bar{e}_x$,
\begin{eqnarray}
& \left(\frac{1}{\sqrt{2}}\right)^* \frac{1}{\sqrt{2}}\left\{-\sum_{k = \{e_x,e_y\}} |\langle k| \bar{e}_x \rangle |^2  +|\langle e_x| \bar{e}_y\rangle | ^2  \right\} & \\ \nonumber
& =  \left(\frac{1}{\sqrt{2}}\right)^* \frac{1}{\sqrt{2}}\left\{- |\langle e_x| \bar{e}_x \rangle |^2  - |\langle e_y| \bar{e}_x \rangle |^2  +|\langle e_x| \bar{e}_x\rangle | ^2  \right\} & \\ \nonumber
& = -1 + 0 - 1 = 0\, ,
\end{eqnarray}
with the terms for ``Case 1'' involving ``$\bar{k}$'' in Eq. \ref{eq:delta_s2} being null.
Similarly, when $i = j = e_y$ and $\bar{a} = \bar{b} = \bar{e}_y$, we also have zero, from just swapping the ``$x$'' and ``$y$'' indices in the above expression.

The pair of ``middle'' terms require ``Case 3,'' where $i \neq j$ and $\bar{a} \neq \bar{b}$:
\begin{eqnarray}
& \left(\frac{1}{\sqrt{2}}\right)^*\frac{1}{\sqrt{2}}\langle e_x | \bar{e}_x \rangle \langle e_y | \bar{e}_y \rangle = \frac{1}{2} ,
 \end{eqnarray}
 and
 \begin{eqnarray}
& \left(\frac{1}{\sqrt{2}}\right)^*\frac{1}{\sqrt{2}}\langle e_y | \bar{e}_y \rangle \langle e_x | \bar{e}_x \rangle = \frac{1}{2} .
\end{eqnarray}

All together, the four contributions give $0+0+1/2+1/2 = 1$. $\Delta \langle \hat{S}^2 \rangle$ is obtained by adding to this $1 - 2M_S^{\textrm{H.S. Ref}} = 1 - 2(1) = -1$. $\Delta \langle \hat{S}^2 \rangle$ for the $^3A_2$ state in this model therefore is then just $1-1 = 0$. Since $\langle \hat{S}^2 \rangle$ for the high-spin reference state was 2, we have, from a simplified version of a minimal description of the SF-BSE computed excitation eigenvector, $\langle ^3A_2 | \, \hat{S}^2 \, |^3A_2\rangle = 2$, as expected.

The reader may verify that a similar approach will give $\langle \hat{S}^2 \rangle = 1$ for either of the $|^3E\rangle$ states, since each is missing a ``partner'' configuration with all of the spins reversed. These configurations are attainable from so-called ``mixed-reference'' spin-flip schemes \cite{lee2018eliminating} but are beyond the scope of present work based on a single reference state.

\section{Execution and timing information of implementation with BerkeleyGW}
\label{sec:timing}

The calculation of $\langle \hat{S}^2 \rangle$ requires the following information: the number of occupied (spin-up) orbitals used in the SF-BSE calculation; the number of unoccupied (spin-down) orbitals used in the SF-BSE calculation; the real, imaginary, and modulus-squared of the overlaps of all of the spin-up occupied orbitals with the same number of spin-down occupied orbitals as well as all of the specified spin-down unoccupied orbitals; and the excitation eigenvector coefficients in the basis of single spin-flip transitions. 

In practice, the overlaps are calculated from files with potentially many more orbitals (or bands) than specified for the SF-BSE calculation; these total number of bands and their occupation values are then also needed. From this latter requirement, we read as input the wavefunction information calculated for the high-spin reference state, in which all unpaired electrons are in the spin-up channel. The wavefunction information, as currently implemented, is provided from an HDF5-format \cite{HDF5} file, \texttt{wfn.h5}, obtained through BerkeleyGW with HDF5 I/O enabled. 
In particular, the data ``\texttt{mnband}'',``\texttt{ifmin}'', and ``\texttt{ifmax}'' from the \texttt{wfn.h5} file are used. As per the BerkeleyGW documentation \cite{berkeleygw_usermanual}, ``\texttt{mnband}'', the total number of bands, is a single integer. ``\texttt{ifmin}'' and ``\texttt{ifmax}'' are integer arrays with two indices; the first (the fast index (first of Python/C, last for Fortran) is the spin index, and the second is the $k$-point index. With the possible exception of defects in two-dimensional materials \cite{arabi2023,thomas2023substitutional}, only the $\Gamma$-point in the Brillouin zone is needed.
``\texttt{ifmin}'' is the lowest occupied state, while ``\texttt{ifmax}'' is the highest occupied state (resolved by spin and, if needed, $k$-point index). The combination of the data from
``\texttt{mnband}'', ``\texttt{ifmin}'', and ``\texttt{ifmax}'' is used to designate orbitals/bands in either spin channel as occupied or unoccupied.

The overlap matrices of the spin-up orbitals with the spin-down orbitals
are computed with the BerkeleyGW utility \texttt{wfn\_dotproduct.x}. The \texttt{overlaps.dat} file contains the necessary overlap data between the spin-up and spin-down orbitals, e.g., $\langle r | \bar{p} \rangle$, that the calculation of $\langle \hat{S}^2 \rangle$ requires (see Eq. \ref{eq:delta_s2}). This is calculated by swapping the spin indices from the wavefunction provided by \texttt{wfn.h5}, copied to a new file \texttt{wfn\_swap.h5}. In a directory with \texttt{wfn.h5} (or a link to it), the following command outputs a file, \texttt{wfn\_swap.h5}:
\begin{displayquote}
\texttt{python spin\_swap.py}
\end{displayquote}
The \texttt{wfn\_swap.h5} file is a copy of the \texttt{wfn.h5} file but with the spin channels reversed; its first spin channel (usually spin-up) is the spin-down channel of \texttt{wfn.h5}, and vice-versa. Using the BerkeleyGW utility \texttt{hdf2wfn.x}, the \texttt{wfn.h5} and \texttt{wfn\_swap.h5} are converted into their corresponding WFN and WFN\_swap files. The \texttt{overlaps.dat} file is computed from the BerkeleyGW executable \texttt{wfn\_dotproduct.x}, with the two input wavefunctions, WFN and WFN\_swap. The output of the \texttt{wfn\_dotproduct.x} execution is piped to a file called \texttt{overlaps.dat}.
The \texttt{eigenvectors.h5} file is generated via a SF-BSE calculation; the protocol for this can be found in Ref. \cite{barker2022spin}. The calculation of $\langle \hat{S}^2 \rangle$ for the high-spin reference state with the Python script \texttt{ref\_s\_sq.py} does not strictly require the \texttt{eigenvectors.h5} file, though this calculation is often performed just before the calculation of  $\Delta\langle \hat{S}^2 \rangle$ in the same directory.

Once the \texttt{wfn.h5}, \texttt{overlaps.dat}, and \texttt{eigenvectors.h5} are generated and/or linked to the desired working directory, the calculations of $\Delta \langle \hat{S}^2 \rangle$ can commence.
The Python script \texttt{ref\_s\_sq.py} is first executed with the command
\begin{displayquote}
\texttt{python ref\_s\_sq.py --wfn \{wfn\_name.h5 \} --nv }\{No. occupied up-spin orbitals\} \texttt{--nc} \{No. unoccupied down-spin orbitals\}.
\end{displayquote}
The output can be piped to a text file, with the last line providing the calculated $\langle \hat{S}^2 \rangle$ for the high-spin reference state. The HDF5-format wavefunction file can have a name different from \texttt{wfn.h5} and is specified by the \texttt{--wfn} directive. 
The Python script \texttt{delta\_s\_sq.py} is then executed with the command
\begin{displayquote}
\texttt{python delta\_s\_sq.py --wfn \{wfn\_name.h5 \} --nv} \{No. occupied up-spin orbitals\} \texttt{--nc} \{No. unoccupied down-spin orbitals\}.
\end{displayquote}
The output of this script should be piped to an output text file. The scripts to calculate $\langle \hat{S}^2 \rangle$ are available publicly in Ref. \cite{SF_BSE_tools_2024}.

We will outline the most important subroutines of the \texttt{delta\_s\_sq.py} script. The subroutine ``\texttt{determine\_case}'' reads in the orbitals involved in the spin-flip transition for the ``bra'' and ``ket'' states and returns the particular case needed. (See Section \ref{sec:example} for a simple illustration.) The subroutine ``\texttt{occ\_ref}'' gives lists for the set of occupied orbitals/bands from the high-spin reference state, as well as the difference of the number of spin-up and spin-down electrons. The subroutine ``\texttt{occupied\_alpha}'' gives a list for the set of the occupied spin-up orbitals/bands for a given spin-flip transition; likewise, the subroutine ``\texttt{occupied\_beta}'' gives a list for the set of the occupied spin-down orbitals/bands for that spin-flip transition. The subroutines ``\texttt{case\_one}'', ``\texttt{case\_two\_a}'', ``\texttt{case\_two\_b}'', and ``\texttt{case\_three}'' calculate, for the particular case, the relevant portion in the curly braces of Eq. \ref{eq:delta_s2}.

By default, \texttt{delta\_s\_sq.py} calculates $\Delta \langle \hat{S}^2 \rangle$ for all excitations. Additionally, for each excitation, the calculation of $\Delta \langle \hat{S}^2 \rangle$ requires loops over each transition-pair $i,\bar{a}$ and $j,\bar{b}$. This gives the script a scaling of $O(N_v^4 N_c^3)$, with $N_v$ the number of specified occupied up-spin orbitals and $N_c$ the number of specified unoccupied down-spin orbitals. The additional factor of $N_v$ comes from the sums over ``$k$'' and/or ``$\bar{k}$'' that occur for all cases except when $i \neq j$ and $\bar{a} \neq \bar{b}$ for a particular pair of transitions. (When $N_v$ is small enough such that it includes only the open-shell spin-up states, these sums are not used, and the scaling is $O(N_v^3 N_c^3)$. These results will not likely be converged, however.) For comparison, the scaling of the diagonalization step of SF-BSE is $O(N_v^3 N_c^3)$. In practice, the pre-factors are typically such that the $\Delta \langle \hat{S}^2 \rangle$ calculation will be only a small fraction of the diagonalization time. The $\Delta \langle \hat{S}^2 \rangle$ scaling could be reduced via an appropriate modification of the script to $O(N_v^3 N_c^2)$, less than that of diagonalization, to compute $\Delta \langle \hat{S}^2 \rangle$ for only a subset of excitations of interest.

\begin{figure}
\centering
\includegraphics[scale=0.8]{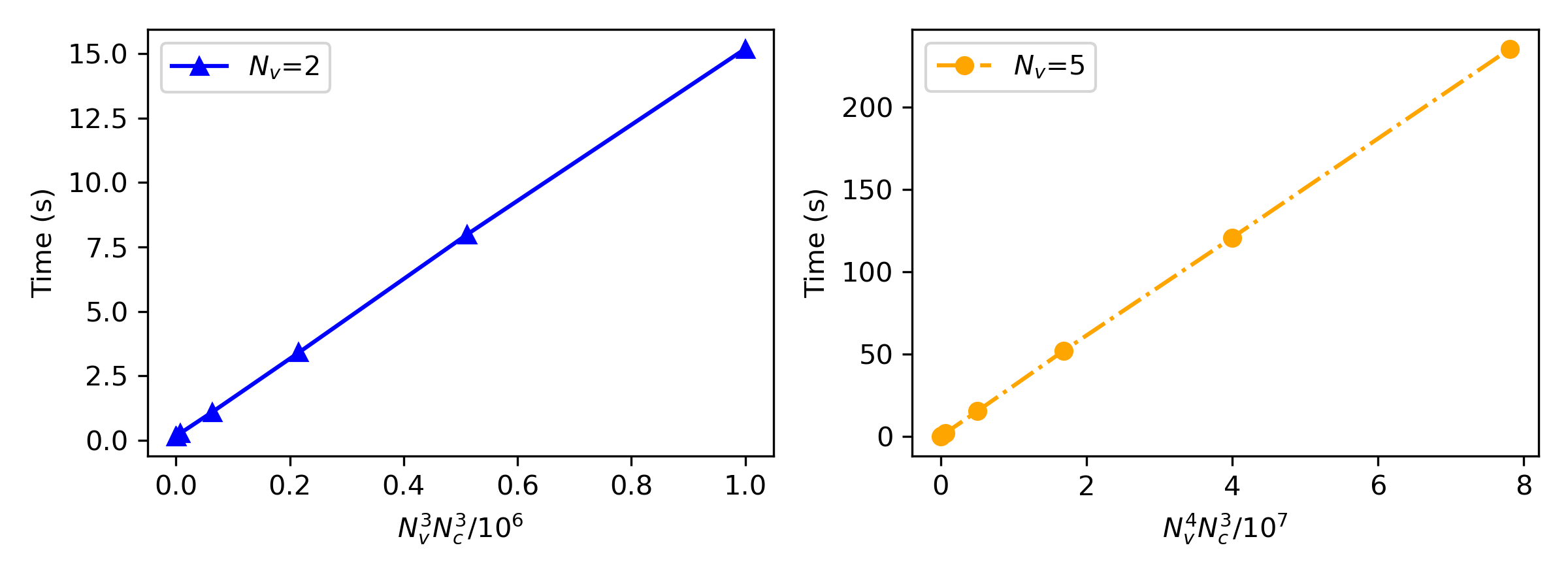}
\caption{Timing information for $\Delta \langle \hat{S}^2 \rangle$ calculations with various sizes of the SF-BSE Hamiltonian for the test system of ethylene under zero torsion. $N_v$ is the number of specified occupied orbitals and $N_c$ is the number of specified unoccupied orbitals. The size of the SF-BSE Hamiltonian is $N_v N_c$. 
For both choices of $N_v$, $N_c$ values from 2 to 50 are considered.}
\label{fig:user_time}
\end{figure}

Timing information is provided for the example case of the ethylene molecule under zero torsion, shown in Fig. \ref{fig:user_time}, with $\Delta \langle \hat{S}^2 \rangle$ calculated for all possible excitations.  The time taken for computations, in seconds, was found by using the \texttt{time} command while running the Python script. In this case, two occupied orbitals are the minimum, and five occupied orbitals are the maximum (limited by the number of occupied down-spin orbitals in the high-reference state).  The upper-limit of 50 empty orbitals is approximately the number of empty orbitals required to converge the ethylene torsion barrier \cite{barker2022spin}. For the calculations with $N_v = 2$, the sums over $k$ and $\bar{k}$ are not performed in Eq. \ref{eq:delta_s2}, and the scaling is $O(N_v^3 N_c^3)$, with a prefactor of  $1.5 \times 10^{-5}\ {\rm s}$. For $N_c = 5$, the scaling is $O(N_v^4 N_c^3)$, with a prefactor of $3.1 \times 10^{-6}\ {\rm s}$ (the same as the $N_v = 2$ case, after accounting for the additional factor of $N_v = 5$.) 

As a test for the convergence of $\Delta \langle \hat{S}^2 \rangle$ (after already obtaining converged values of the excitation energies from SF-BSE), we compare the values for the cases ($N_v = 5$, $N_c = 5$) with ($N_v = 5$, $N_c = 50$), and find differences, at most, on the order of 0.01. (See Table \ref{tab:s2_convergence}.) Therefore, one may take $\Delta \langle \hat{S}^2 \rangle$ to be computed sufficiently with fewer unoccupied states than were used to construct the SF-BSE Hamiltonian.

\begin{table}[ht]
    \label{tab:s2_convergence}
    \caption{The computed $\Delta \langle \hat{S}^2 \rangle$ values for the states N, T, V, and Z for ethylene under zero torsion, with two different sizes of the orbital basis set. The last row is the expected values for pure singlets (-2) and triplets (0).}
    \centering
    \begin{tabular}{|c|c|c|c|c|c|}
        \hline
        $N_v$ & $N_c$  & N & T & V & Z \\
        \hline \hline
         5 & 5  & -1.985 & -0.045 & -1.965 & -1.957 \\
         5 & 50 & -1.998 & -0.014 & -1.974 & -1.983 \\ \hline
         - & - & -2 & 0 & -2 & -2 \\
        \hline
    \end{tabular}
\end{table}

\section{Conclusion}
\label{sec:conclusion}

We showed the necessity of the computation of $\langle \hat{S}^2 \rangle$ to make meaning of the results from Spin-Flip Bethe-Salpeter Equation calculations due to ambiguities from phase differences in orbitals in different spin channels, and ease of rapid identification of desired excited states. We reviewed the notion of spin contamination and its possible sources from spin-polarized calculations (generally giving small contamination) and incomplete transition vectors (possibly giving large contamination). We then derived in great detail the equation used for calculating $\Delta \langle \hat{S}^2 \rangle$ from the ``super-operator'' approach, by explicitly considering the four cases possible for the individual transitions in the bra and ket states. We reviewed the suite of Python scripts written for BerkeleyGW to calculate $\langle \hat{S}^2 \rangle_0$ for the high-spin reference state and $\Delta \langle \hat{S}^2 \rangle$ for the excited states, so that this approach may be implemented independently for research teams interested in the spin-flip method. These scripts will be available in a future release of the BerkeleyGW package. We included timing information for the execution of the (serial) script for $\Delta \langle \hat{S}^2 \rangle$, showing $O(N_v^4N_c^3)$ scaling, when computing for all possible excitations. The convergence of $\Delta \langle \hat{S}^2 \rangle$ for our two examples of ethylene and the NV$^{-}$ center in diamond indicated rapid convergence with respect to the number of unoccupied states, so the serial script with fewer unoccupied states may be performed initially for the purposes of analysis of states before the full set of unoccupied states are used for fully-converged results for $\Delta \langle \hat{S}^2 \rangle$ values.

\ack{The authors thank Hrant Hratchian and Abdulrahman Zamani for useful discussions about spin-contamination. 
B.A.B. and D.A.S. were supported by the U.S. Department of Energy, Office of Science, Basic Energy Sciences, CTC and CPIMS Programs, under Award DE-SC0019053. A.S. and D.A.S. were supported by the U.S. National Science Foundation under Grant No. DMR-2144317. A.S. was additionally supported by a fellowship from the Achievement Rewards for College Scientists (ARCS) Foundation, Northern California Chapter.
Computational resources were provided by the National Energy Research Scientific Computing Center (NERSC), a U.S. Department of Energy Office of Science User Facility operated under Contract No. DE-AC02-05CH11231, and by the Multi-Environment Computer for Exploration and Discovery (MERCED) cluster at UC Merced, funded by National Science Foundation Grant No. ACI-1429783.}

\bibliographystyle{iopart-num}
\bibliography{references}

\end{document}